# Enhancement of the electron spin memory by localization on donors in a quantum well

J. Tribollet[1], E. Aubry[1], G. Karczewski[2], B. Sermage[3], F. Bernardot[1], C. Testelin[1] and M. Chamarro[1]

(1) Institut des NanoSciences de Paris - Universités Paris-VI et Paris-VII, CNRS UMR 7588, 2 place Jussieu, 75251 Paris cedex 05, France
(2) Institute of Physics, Polish Academy of Sciences, Al. Lotnikow 32/46, 02-668 Warsaw, Poland.
(3) CNRS-Laboratoire de Photonique et Nanostructures, Route de Nozay, 91460 Marcoussis, France

**ABSTRACT**

We present easily reproducible experimental conditions giving long electron spin relaxation and dephasing times at low temperature in a quantum well. The proposed system consists in an electron localized by a donor potential, and immerged in a quantum well in order to improve its localization with respect to donor in bulk. We have measured, by using photoinduced Faraday rotation technique, the spin relaxation and dephasing times of electrons localized on donors placed in the middle of a 80Å CdTe quantum well, and we have obtained 15ns and 18ns, respectively, which are almost two orders of magnitude longer than the free electron spin relaxation and dephasing times obtained previously in a similar CdTe quantum well (J. Tribollet *et al.* PRB 68, 235316 (2003)).

**INTRODUCTION**

In last years, interest in spin physics has been renewed due to its potential application in spintronics and quantum information [1]. In this framework, the main required property is the presence of long spin memory. Several studies in III-V bulk n-doped semiconductors performed in the past demonstrated a long electron-spin relaxation time at low temperature [2-4]. That is why the electron spin is a promising candidate to form a "qubit" in future quantum computers. Recent studies of the spin lifetime as a function of donor concentration, in n-doped GaAs crystals, have revealed a maximum of about 100 ns for a concentration near $10^{16} cm^{-3}$, and this behaviour has been explained as an interplay of different relaxation mechanisms [5].

In bulk materials, doping atoms and electrons coexist in the same sample and, at low temperature, electrons are mainly localized on donors. Epitaxy techniques allow to separate electrons from donors. Indeed, donors can be introduced in the material which forms barriers in such a way that electrons migrate to the quantum well (QW) material, creating a 2D free electron gas. There are several studies of spin relaxation in doped QWs [6-13] but, the spin relaxation time is mainly of the order of several hundreds of picoseconds. When a longer lifetime was observed, in general, authors claimed about the localization of electrons in the fluctuations potential of QW, associated to the spatial variations in the density of donors in the barrier [12,13].

In this paper, we present easily reproducible experimental conditions giving long electron spin relaxation and dephasing times at low temperature in a quantum well. We focus our study on electrons localized on donors which have been inserted in the middle of a 80Å QW. This system allows us to increase the localization of electron wavefunction, with respect to localization on donors in 3D crystals [14]. We have chosen, here, a CdTe QW in such a way that a comparison with previous results on free electrons [9] be possible. To the best of our knowledge there is no study of spin relaxation or dephasing times in such a system. We will discuss advantages and disadvantages, when a long spin memory is wanted, of CdTe compound *versus* the most currently studied GaAs compound.

We use a pump-probe technique, the photoinduced Faraday rotation (PFR), which is well adapted for studying spin relaxation and dephasing times of resident electrons. By comparing the experimental results with previous results obtained for free electrons in a similar CdTe QW, we show that the localization of electrons enhances its spin memory by almost two orders of magnitude.

**SAMPLE CHARACTERISATION**

The studied sample consists in a CdTe/CdMgTe heterostructure grown by molecular-beam epitaxy on (100)-oriented GaAs substrate, and containing a 80Å QW. A donor layer of iodine was placed in the center of the QW. Donor concentration is approximately $10^{11}$ cm$^{-2}$. In order to perform transmission and PFR measurements we have chemically suppressed the GaAs substrate.

Figure 1 shows low-temperature transmission and luminescence spectra of the studied sample. Transmission spectrum is dominated by a broad band with minimum at 1.6195 eV, and a shoulder at lower energy 1.615 eV. Kheng *et al*. [15] have shown that the introduction of donors in a CdTe QW leads to the observation, in absorption spectra, of a band associated to the formation of an exciton bound to a neutral donor, called $D^0X$. Vertical arrow indicates the energy of the transmission minimum of a 80 Å QW without doping layer, i. e., the energy necessary to form a free exciton (1.624 eV). The energy difference between the maxima of an empty QW and a doped QW, allows us to determine the binding energy of the exciton bound to a neutral donor $D^0X$, *i. e.* 4.5 meV. The shoulder at lower energy is assigned to the formation of an exciton bound to a neutral acceptor. Indeed, the introduction of donor impurities in the QW creates compensation sites, then follows the presence of acceptor sites which, in our case, are probably cadmium vacancies. The luminescence spectrum, obtained after excitation with a 5 mW 633nm He-Ne laser, is also shown in figure 1. It is dominated by the recombination of excitons bound to acceptors, with a long tail at low energy and a very small shoulder at high energy corresponding to the recombination of excitons bound to donors.

Picosecond pulses from a Ti:sapphire laser are used to excite the photoluminescence of $D^0X$, while a streak camera placed after a double monochromator records its time-resolved decay. The insert in figure 1 shows, in particular, the time-resolved PL obtained under resonant excitation of $D^0X$ (1.619 eV). Due to this resonant excitation the first 40 ps are affected by laser diffusion, even when a linearly polarized light was used in the excitation and the cross-polarized emission was detected. Decay signal has a mono-exponential behaviour with a characteristic time of 175 ps. A similar single exponential decay was also observed for a non-resonant excitation.

**PHOTO-INDUCED FARADAY ROTATION MEASUREMENTS AND DISCUSSION**

The light source of our PFR experimental set-up is a Ti:sapphire laser with a 2 ps pulse duration and a repetition rate of 76MHz, which is split into pump and probe beams. Pump average intensity was in the order of 1 W/cm$^2$, and average probe intensity was ten times less. Pump beam polarization is modulated at 42 KHz by using a photoelastic modulator; probe beam is linearly polarized, and its intensity is modulated with an optical chopper at 1KHz. After transmission through the sample, the rotation angle of probe beam polarization is analysed in an optical bridge [9]. To improve signal-to-noise ratio, a double lock-in amplifier analysis of rotation angle was performed.

### a) Spin relaxation measurements

Figure 2 shows the PFR signal obtained at low temperature, 2K, when pump and probe beams are tuned to 1.619 eV. The band width of the used mode-locked Ti:sapphire is less than 1 meV, allowing the pump to create, mainly, excitons bound to neutral donors. We have to note a non-zero PFR signal at negative pump-probe delay times, indicating that the electron spins are not fully relaxed within the 13 ns repetition period of our Ti:sapphire laser. Assuming an exponential decay of the long-living PFR signal, and taking the extrapolated signal value at t=0 and the value at t<0, we have evaluated the decay time of this long signal to be 15ns. According to the already discussed results of time-resolved photoluminescence, this long-living signal has to be related to the net spin polarization of electrons bound to donors, which are the only species present in the sample when the recombination of $D^0X$ is accomplished. The spin polarization of resident electrons is built *via* the polarization of $D^0X$, resonantly excited by a circularly polarized pump pulse, and the subsequent transfer of this polarization to the electrons. Mechanisms involved in this transfer are very similar to those present in the polarization of electrons of a 2D gas *via* the resonant excitation of trions [9]. The main condition for this spin transfer is that the spin relaxation time of photocreated holes keeps in the same order of magnitude or smaller than the recombination time of $D^0X$.

Insert in figure 2 gives a diagram of different optical transitions and relaxation times involved in the formation and evolution of $D^0X$ complexes. The $\sigma+(\sigma-)$ circularly polarized pump, photocreates a $D^0X$ with a +3/2 (-3/2) hole, from a spin up (down) electron bound to a donor. The corresponding localized complex is denoted $D^0X_{+3/2}$ ($D^0X_{-3/2}$). Immediately after the pump pulse, the sample contains donors with spin-polarized electrons up (down) and also polarized complexes $D^0X$. The spin of a $D^0X$ complex is the spin of the photocreated hole, because its two electrons, one photocreated and the other associated to the initial neutral donor, are in a singlet state in the lowest energy state. $T^h$ is the spin relaxation time of the hole, and $T_1$ is the spin relaxation time of electrons. $T_R$ is the recombination time of $D^0X$. If $T^h \leq T_R << T_1$, during the presence in the sample of $D^0X$ complexes, the populations of $D^0X_{+3/2}$ and $D^0X_{-3/2}$ tend to balance, and the recombination of $D^0X$ complexes gives localized electrons with a polarized spin up or down with almost the same probability. Once the $D^0X$ relaxation achieved, the number of resident electrons having spin down (up) is larger than the number of electrons having spin up (down), and a PFR signal can then be detected.

The PFR decay curve given in figure 2 can be fitted by three exponential decays: a very fast decay with a short constant time of about 10ps, a middle one with an associated constant time of about 80ps, and a very long third-one with a characteristic time of 15ns. The two shorter times are smaller than the $D^0X$ recombination time, and then they have to be associated to the presence of $D^0X$ complexes. The decay of the difference between the $D^0X_{+3/2}$ and $D^0X_{-3/2}$ populations follows a single exponential with a decay rate $1/T=1/T_R+1/T^h$. This rate comes from the two possible channels this population difference diminishes, one being the recombination of $D^0X$, and the second one the hole spin flip at constant total population. Taking T=80ps we deduce a hole spin relaxation time of $T^h$ =147ps, which is slightly smaller but of the same order of magnitude than the experimentally determined recombination time $T_R$=175 ps. However, if we take T=10ps, the associated hole spin relaxation time is equal to $T^h$ = 10 ps clearly shorter than the recombination time. We have no model for the presence in the same sample of these two short times (10 ps and 80 ps), but we remark that assuming a unique recombination time of 175ps, in both cases the condition $T^h \leq T_R$ for the polarization of electrons bound to donors is fulfilled. Thus, we are able to polarize the spin of an electron localized on a donor, and we have determined its characteristic damping time, $T_1$=15ns. This time is larger by two orders of magnitude than the spin relaxation time obtained for free electrons in a 2D gas in a similar CdTe quantum well [9].

**b) Spin dephasing measurements**

We have also measured the PFR signal in presence of a transverse magnetic field. The magnetic field is applied along x direction, and the pump and probe beams are aligned to z direction, which is also the growth direction of the sample. As already explained, at t=0 pump beam creates polarized $D^0X$ complexes and polarized electrons. When a transverse magnetic field is applied in the QW plane, this field induces a precession of the net spin polarization of electrons, but not for $D^0X$. We have already mentioned that a $D^0X$ contributes to the spin dynamics by its hole only. The precession frequency is the Larmor frequency, $\Omega_{e,h} = \dfrac{g^{\perp}_{e,h}\mu_B}{\hbar}B$, and is almost zero for holes because $g^{\perp}_h \approx 0$, which is not the case for electrons. Then the spin dynamics after initialization by a short pump pulse can be described by the following general equation:

$$\frac{d\vec{S}_\perp}{dt} = \vec{\Omega}_e \wedge \vec{S}_\perp - \frac{\vec{S}_\perp}{T_2} + \frac{J(t)}{T_R}\vec{e}_z \qquad (1)$$

where $\vec{S}_\perp$ is the transverse component to the applied magnetic field of the total electronic spin, $\vec{\Omega}_e = \Omega_e \vec{e}_x$, $T_2$ is the decoherence time for a electron and $J(t) = J_0 e^{-\frac{t}{T}}$ is one third of the kinetic momentum of $D^0X$ as a function of time with $J_0$ its value at t=0. Solution of this equation gives a simple expression for $S_z$ when t>>T:

$$S_z(t) \propto e^{-\frac{t}{T_2}} \cos(\Omega_e t) \qquad (2)$$

Figure 3 shows that the signal is dominated by an oscillatory behaviour, which is associated to the polarization of the electronic spins. We have fitted all PFR curves to expression (2). Figure 4 a) shows the dependence of $\Omega_e$ on the applied magnetic field. A linear fit of the data gives an electron Landé factor $|g_e^\perp| = 1.3$, which is comparable to the value obtained for a free electron in a similar CdTe quantum well [9, 11].

For low magnetic fields, less than 0.59 T, the oscillatory behaviour of PFR signal is also observed at negative time delays. That means, firstly, that the damping time of oscillations is comparable to the 13ns which is the time interval of our laser, and secondly, that this damping time decreases for increasing magnetic fields. Figure 4 b) shows the experimentally determined $1/T_2$, denoted $1/T_2^*$, as a function of magnetic field. We have observed, in the field range of 0.082T to 1.18 T, a linear dependence of $1/T_2^*$ on magnetic field, leading to a $T_2^*$ ranging from 9.4 ns to 1.5 ns. The used PFR technique gives information on an ensemble of electronic spins, and it is then sensitive to inhomogeneities, such as local magnetic fields or local variations in the electron g-factor. Assuming a lorentzian distribution of g factors with half width at half maximum $\Delta g_e^\perp$, the experimentally determined $T_2^*$ is related to the $T_2$ which is not affected by g inhomogeneities, as follows:

$$\frac{1}{T_2^*} = \frac{1}{T_2} + \Delta g_e^\perp \frac{\mu_B}{\hbar} B \qquad (3)$$

From a linear interpolation to the expression (3) of the experimentally determined spin dephasing rates $1/T^*_2$, we obtain $T_2 \approx 18$ns and $\Delta g_e^\perp \approx 0.005$. The latter value leads to $\frac{\Delta g_e^\perp}{g_e^\perp} \approx 0.4\%$ which is similar to the value obtained for free electrons in other QWs [11]. The determined $T_2$ value could yet be affected by inhomogeneities in the sample. In bulk materials, it is well established that below Mott transition there are two main mechanisms responsible for the electronic spin relaxation; their relative importance depends on donor concentration: the hyperfine interaction is important at low donor concentration and the anisotropic exchange interaction dominates at higher donor concentration. Donor environments could be different from one donor to another and then could induce different values of spin decoherence time.

It is worth noting that the spin coherence time of a neutral donor is longer than the spin coherence time of 2.5ns reported for localized trions in GaAs [12], and shorter than 100 ns reported in bulk GaAs [4,5]. At this point, a comparison of the importance of different spin relaxation mechanisms in GaAs and CdTe is important. According to the previously referenced theoretical works [5], at high donor concentrations, the strength of anisotropic exchange interaction is proportional to $\alpha^2$, with $\alpha$ a dimensionless factor appearing at 3erd power in the k term os the conduction-band Hamiltonian (see reference [16]). This factor has been experimentally determined for GaAs, but no experimental data can be found for CdTe. In order to obtain an $\alpha$ estimat for both compounds, we have used the following expression [17]:

$$\alpha \approx \frac{4\Delta_{SO}}{\sqrt{(E_g + \Delta_{SO})(3E_g + 2\Delta_{SO})}} \frac{m^*}{m_0} \quad (4)$$

where $\Delta_{SO}$ is the energy difference between the valence band ($\Gamma_8$) and spin-orbit band ($\Gamma_7$), $E_g$ is the energy gap of the compound, and m* is the electron effective mass. Taking values from ref. [18] for GaAs and from ref. [19] for CdTe, we obtain that the anisotropic interaction is stronger in CdTe than in GaAs by a factor of nine leading to shorter spin lifetimes in CdTe.

As already mentioned, the most important spin relaxation mechanism for isolated localized electrons in semiconductors is the hyperfine interaction. Spin dephasing time for an ensemble of donors is fixed by the value of the nucleus hyperfine constant, A, the value of nuclear spin momentum, I, and the number of atoms $N_L$ effectively seen by an electron, as follows [20]:

$$\frac{1}{T_2} \propto \sqrt{\frac{\sum_i I^i(I^i+1)(A^i)^2}{N_L}} \quad (5)$$

where the sum is over different nuclei in the material. There are no experimental determination of A constants for CdTe, but theoretical estimations give an A constant for $Cd^{111}$ almost four or five times smaller than the same constant for Ga and As atoms [20]. Wavefunction localization is stronger in II-VI compounds compared to in III-V compounds. Moreover, in a CdTe sample only one sixth of the total amount of atoms have a nucleus carrying a non-zero nuclear spin momentum (equal to one half), whereas in more common III-V compounds (GaAs, AlAs, InAs,….) all atoms have nuclei carrying a higher nuclear spin momentum. Finally, taking into account all the value of all these parameters, hyperfine interaction is of the same order of magnitude or slightly weaker in CdTe than in more commonly studied III-V semiconductors, allowing us to expect longer spin relaxation or dephasing times in CdTe, after optimization of experimental parameters of the sample.

**CONCLUSION**

In conclusion, we have obtained the enhancement, by two orders of magnitude, of the spin memory of electrons when they are localized by the coulombic potential of donors immerged in QW. The proposed system seems very interesting when we compare the value of spin dephasing time that we have obtained (18 ns) with the one obtained recently for electrons localized on GaAs [21] and InAs [22] quantum dots, which are of the order of 10ns. Thus, electrons localized on donors are a reproducible and model system of n-doped quantum dots giving the possibility to study the electronic spin dephasing time as a function of the electron localization controlled by the spatial position of donors in QW and/or by controlling the QW thickness.

Authors acknowledge financial support of the Research Ministry through an "ACI Jeunes Chercheurs 2002" grant, and by the Île-de-France Regional Council through the "projet SESAME 2003" n° E.1751.

# REFERENCES

[1] "Semiconductor Spintronics and Quantum Computing" ed by D.D. Awschalom, D. Loss and N. Samarth (Springer, New York, 2002).

[2] C. Weisbuch, Ph D. thesis, University of Paris. (1997).

[3] R.I. Dzhioev, B.P. Zakharchenya, V.L. Korenev, and M.N. Stepanova, Phys. Solid State 39, 1768 (1997).

[4] J.M. Kikkawa and D.D. Awschalom, Phys. Rev. Lett 80, 4313, (1998).

[5] R.I. Dzhioev, K.V. Kavokin, V.L. Korenev, M.V. Lazarev, B. Ya. Meltser, M.N. Stepanova, B.P. Zakharchenya, D. Gammon, and D.S. Katzer, Phys. Rev. B 66, 24524, (2002).

[6] T.C. Damen, L. Vina, J.E. Cunningham, and J. Shah, Phys. Rev. Lett. 67, 3432, (1991).

[7] J.M. Kikkawa, I.P. Smorchkova, N. Samarth, and D.D. Awschalom, Science 277, 1284, (1997).

[8] J.S. Sandh, A.P. Heberle, J.J. Bamberg, and J.R.A. Cleaver, Phys. Rev. Lett. 86, 2150, (2001).

[9] J. Tribollet, F. Bernardot, M. Menant, G. Karczewski, C.Testelin, and M. Chamarro Phys. Rev B 68, 235316 (2003).

[10] Y. Ohno, R. Terauchi, T. Adachi, F. Matsukura and H. Ohno, Phys. Rev. Lett. 83,4196, (1999).

[11] E.A. Zhukov, D. R. Yakovlev, M. Bayer, G. Karczewski, T. Wojtowicz, and J. Kossut, Phys. Stat. Sol. (b) 243, 878, (2006). R. Bratschitsch, Z. Chen, S.T. Cundiff, D.R. Yakovlev, G. Karczwski, T. Wojtowicz and J. Kossut. Phys. Stat. Sol. (b) 243, 2290 (2006).

[12] T.A. Kennedy, A. Shabaev, M. Scheibner, Al.L. Efros, A. S. Brackar and D. Gammon, Phys. Rev. B73, 04307 (2006).

[13] H. Hoffmann, G. V. Astakhov, T. Kiessling, W. Ossau, G. Karczewski, T. Wojtowicz, J. Kossut, and L.W. Molenkamp. Phys. Rev. B 74, 073407 (2006)

[14] P. Harrison, S. J. Weston, T. Piorek, T. Stirner, W.E. Hagston, J.E. Nicholls and M. O'Neill Superlattices and Microstructures, 14, 249 (1993).

[15] K. Kheng, R.T. Cox, Y. Merle d'Aubigné, F. Bassani, K. Saminadayar, and S. Tatarenko, Phys. Rev. Lett. 71, 1752 (1993). K. Kheng, Ph. D. thesis, Grenoble, (1995).

[16] "Optical orientation" edited by F. Meier and B. Zakharchenya vol8 of Modern Problems in Condensed Matter Sciences (Noth-Holland, Amsterdam, 1984).

[17] M. I. D'yakonov and V. Yu. Kachorovskii, Soviet Physics Semiconductors 20, 110, (1986).

[18] U. Rössler, Solid State Comm. 49, 943,(1984).

[19] H. Mayer and U Rössler. Solid State Comm. 87, 81(1993).

[20] I.A. Merkulov, Al.L. Efros, and M. Rosen PRB 65, 205309 (2002).

[21] M. V. Gurudev Dutt, Jun Chebg, Bo Li, Xiaodong Xu, Xiaoqin Li, P.R. Berman, D. G. Steel, A. S. Bracker, D. Gammon, S.E. Economou, Ren-Bao Liu, and L.J. Sham, Phys.Rev Lett 94, 227403 (2005).

[22] A. Greilich, R. Oulton, E. A. Zhukov, I. A. Yugova, D. R. Yakovlev, M. Bayer, A. Shabaev, Al.L. Efros, A.I. Merkulov, V. Stavarache, D. Reuter, and A. Wieck, Phys. Rev. Lett. 96, 227401 (2006).

**FIGURE CAPTIONS**

**Figure 1**. Transmission (full line) and photoluminescence (dashed line) spectra obtained at 2K for a 80Å CdTe QW with a donor layer of iodine atoms placed in its center. Vertical arrow indicates the energy of the minimum of transmission observed for a 80 Å CdTe QW without doping layer. Insert represents the 10K photoluminescence decay observed at 1.619 eV after resonant excitation. Black line corresponds to an exponential fit with a characteristic time 175 ps.

**Figure 2**. PFR signal as a function of pump-probe delay, obtained at 2 K for degenerated pump-probe beams tuned to 1.619 eV. Full line represents a fit to a three exponential decay with characteristic times 10ps, 80ps and 15ns. Insert shows a diagram of different optical transitions and relaxation times involved in the formation and evolution of $D^0X$ complexes.

**Figure 3** PFR signal as a function of pump-probe delay obtained at 2K for several values of a transverse magnetic field. Curves have been shifted for clarity. Dashed lines fixe the zero of each curve.

**Figure 4 a)** Larmor frequency *versus* applied transverse magnetic field. A linear fit gives the Landé factor $|g_e^\perp|$=1.3; **b)** Inverse of the spin dephasing time *versus* applied transverse magnetic field. A linear fit according to equation (3), gives the dephasing time free from g inhomogeneities, and $\Delta g$ (see the text).

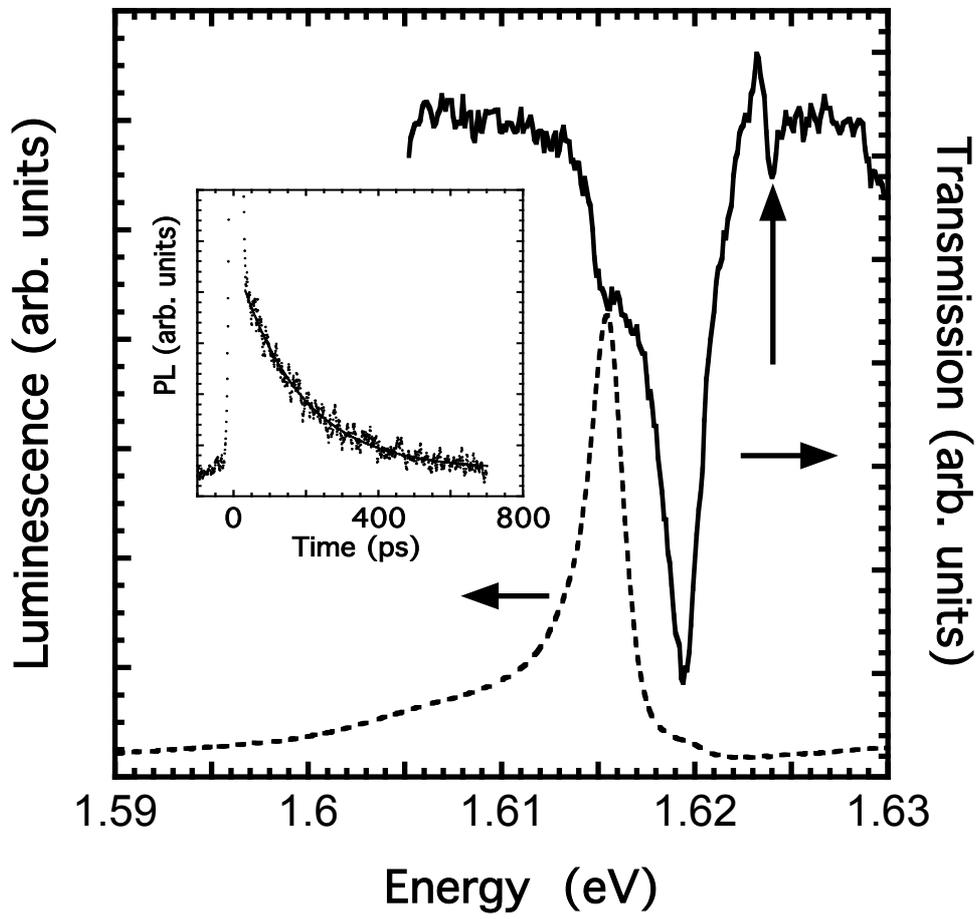

Figure 1

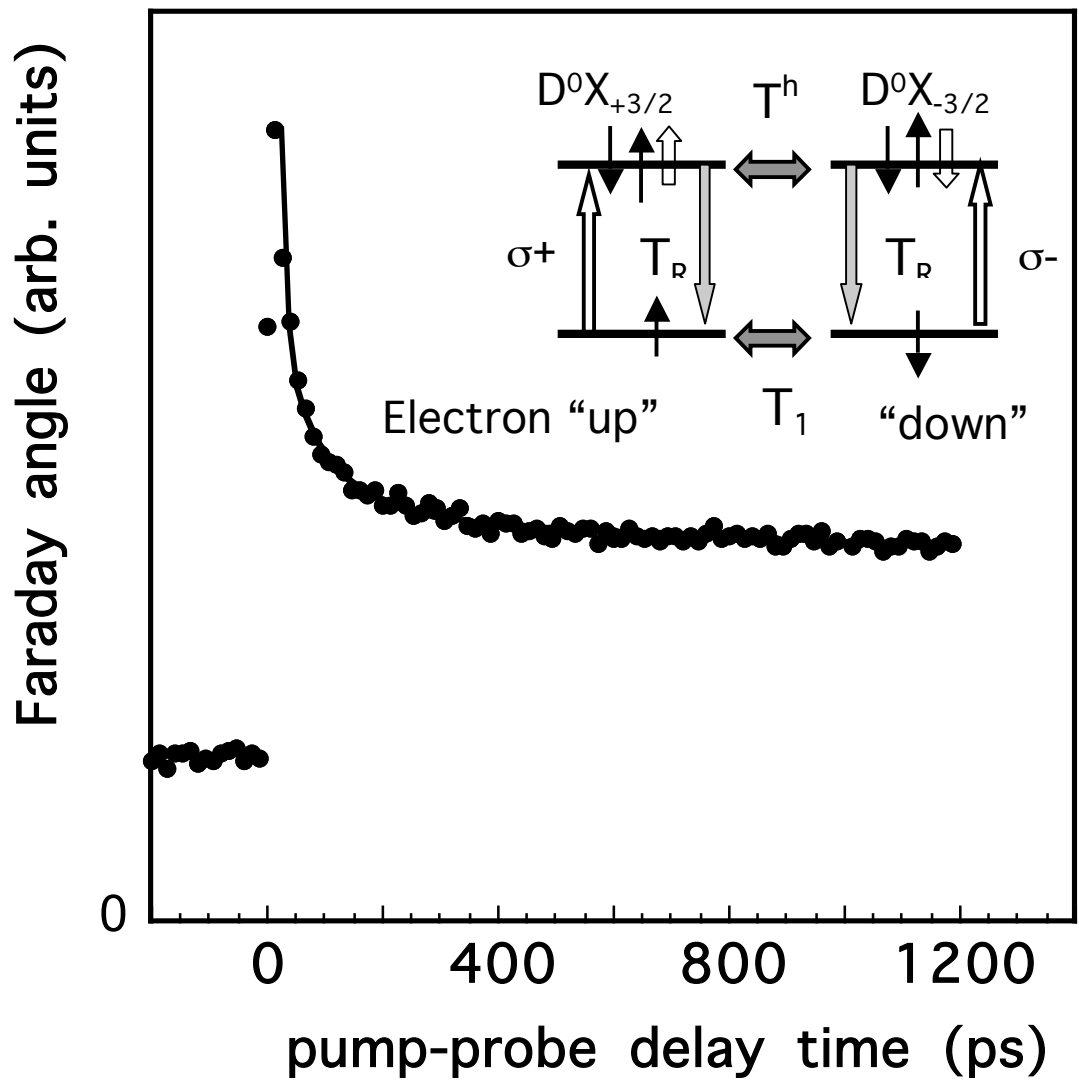

Figure 2

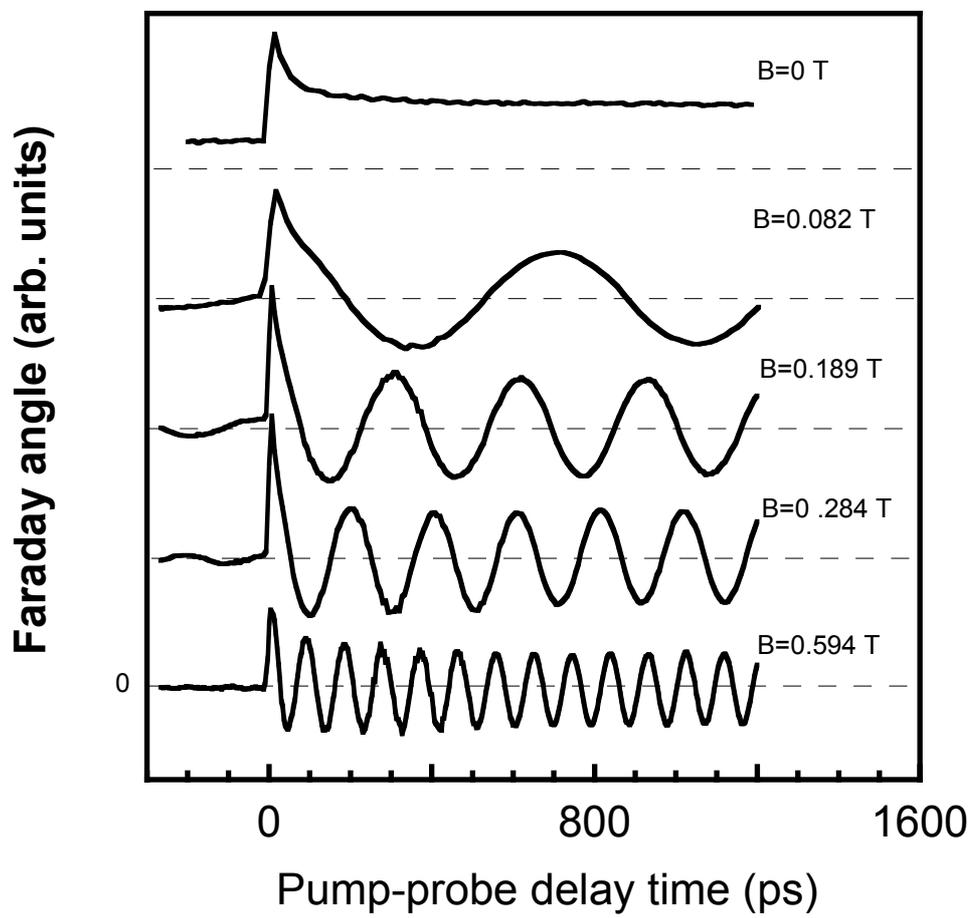

Figure 3

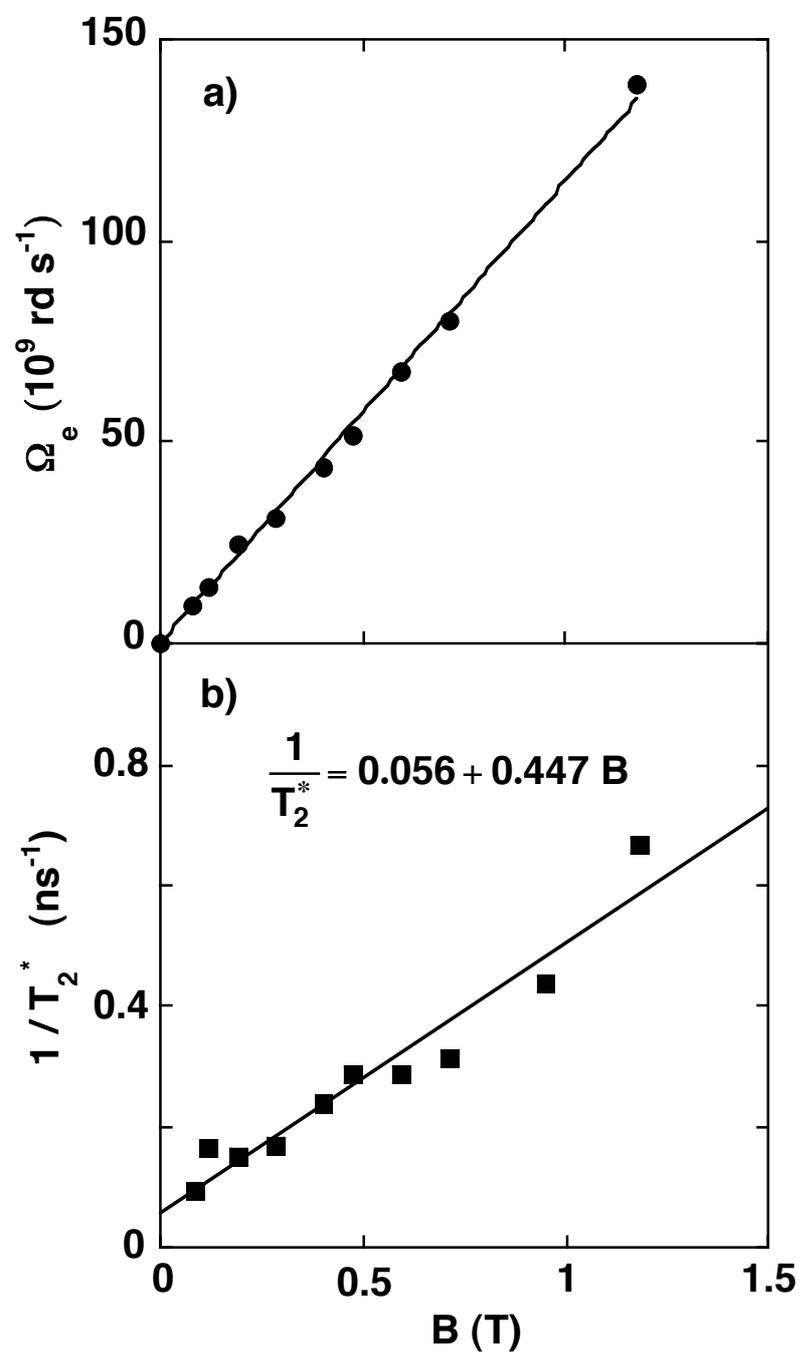

Figure 4